\documentstyle[twoside,fleqn,espcrc2]{article}

\begin{document}

\newcommand{\ttbs}{\char'134}
\newcommand{\AmS}{{\protect\the\textfont2
  A\kern-.1667em\lower.5ex\hbox{M}\kern-.125emS}}

\title{ADM Worldvolume Geometry}
\author{R. Capovilla\address{F\'{\i}sica, CINVESTAV-IPN,
Apdo. Postal 14-740, 07000 M\'exico D.F., M\'exico}\thanks{
Supported in part by the organizers and by the Physics Department of
CINVESTAV-IPN. }, J. Guven\address{Instituto de Ciencias Nucleares,
 Universidad Nacional Aut\'onoma de M\'exico, 
 Apdo. Postal 70-543, 04510 M\'exico D.F., M\'exico}
 and E. Rojas$^{\rm a}$}
\begin{abstract}
We describe the dynamics of a relativistic extended object in terms of the
geometry of a configuration of constant time. This involves an adaptation of
the ADM formulation of canonical general relativity. We apply the formalism
to the hamiltonian formulation of a Dirac-Nambu-Goto [DNG] relativistic 
extended object in an arbitrary background spacetime.
\end{abstract}

\maketitle

The mechanics of relativistic extended objects, or branes 
for short, has
been the subject of various investigations in recent years. 
In particular, the 
usefulness of a geometrical approach which mantains general 
covariance with
respect to worldvolume reparametrizations and rotations of worldvolume
normal vector fields has become apparent (see {\it e.g.} \cite{geo1,geo2}). 
However, this geometrical
approach has been developed largely within the lagrangian formulation of
theories of relativistic extended objects. The geometry of interest is then
the worldvolume geometry. In this note,  we report on an alternative 
approach where we focus on the geometry of the relativistic extended object 
itself, and its relation with the worldvolume geometry \cite{CGR1}. 
One immediate application of this formalism is a geometrical 
Hamiltonian formulation of this type of theories, which mantains 
worldvolume covariance without resorting to gauge fixing 
from the outset. (For alternative treatments see
{\it e.g.} \cite{CHT,Hoppecan,Smolin}).

We consider a relativistic extended object $\Sigma $, of dimension $d$,
embedded in an arbitrary fixed $(N+1)$-dimensional background spacetime $
\{M,g_{\mu \nu }\}$. $\Sigma $ is described locally by the spacelike
embedding $x^{\mu }=X^{\mu }(u^{A})$, where $x^{\mu }$ are local coordinates
for the background spacetime, $u^{A}$ local coordinates for $\Sigma $, and $
X^{\mu }$ the embedding functions ($\mu ,\nu ,\cdots =0,1,\cdots ,N$, and $
A,B,\cdots =1,\cdots ,d$). The tangent vectors to $\Sigma $ are defined
by $\epsilon _{A}:=(\partial _{A}X^{\mu })\partial _{\mu }$, so that the
positive-definite metric induced on $\Sigma $ is 
\begin{equation}
h_{AB}:=g(\epsilon _{A},\epsilon _{B})\,.
\end{equation}
We construct out of the metric $h_{AB}$ the intrinsic geometry of $\Sigma $.
Note that it will be trivial in the special case of a relativistic string,
since then $\Sigma $ is one-dimensional.

For the extrinsic geometry of $\Sigma $, we introduce the normals $\{m^{I}\}$
, defined by  $g(\epsilon _{A},m^{I})=0$, and normalized with $
g(m^{I},m^{J})=\eta ^{IJ}$, where $\eta _{IJ}$ is the Minkowski metric, with
only one minus sign ($I,J,\cdots =0,1,\cdots ,N+1-d$). The extrinsic
curvature along the $I$-th normal is 
\begin{equation}
L_{AB}{}^{I}:=-g(m^{I},D_{A}\epsilon _{B})=L_{BA}{}^{I}\,,
\end{equation}
where $D_{A}=\epsilon ^{\mu }{}_{A}\,D_{\mu }$ is the spacetime covariant
derivative compatible with $g_{\mu \nu }$ along the tangent directions. In
addition, the extrinsic geometry of $\Sigma $ is determined by the extrinsic
twist, $\sigma _{A}{}^{IJ}$, defined by 
\begin{equation}
\sigma _{A}{}^{IJ}:=g(m^{J},D_{A}m^{I})=-\sigma _{A}{}^{JI}\,.
\end{equation}
This is a connection associated with the $O(N+1-d)$ freedom in the
definition of the normal fields. In the mathematical literature it is known
as the normal form. When the appropriate Gauss-Codazzi-Mainardi integrability
conditions hold, $\{h_{AB},L_{AB}{}^{I},\sigma _{A}{}^{IJ}\}$ define the
geometry of $\Sigma $.

We consider now the time evolution of $\Sigma $ in spacetime. We denote its
trajectory, or worldvolume, by $w$. It is an oriented timelike surface
in
spacetime. Now the shape functions become time-dependent,
$X^\mu = X^\mu ( t , u^A )$, where $t$ is a coordinate that labels
the leafs of the foliation of $w$ by $\Sigma$s.

From the point of view of an observer sitting on $\Sigma $, this
involves breaking the normal rotation symmetry of $\Sigma $, $O(N+1-d)$,
down to $O(N-d)$, by choosing the unit (future-pointing) timelike normal to $%
\Sigma $ into $w$, $m^{0}=:\eta $. We denote by $m^{i}=:n^{i}$ the
remaining components of $\{m^{I}\}$ ($i,j,\cdots =1,2,\cdots ,N-d$). These
are also normals to the worldvolume $w$. This symmetry reduction is
the key ingredient in this geometrical construction.

The time evolution of the embedding functions for $\Sigma $ into the
worldvolume can be written as, 
\begin{equation}
\dot{X}^{\mu }:=N\eta ^{\mu }+N^{A}\epsilon ^{\mu }{}_{A}\,, 
\label{eq:vel}
\end{equation}
where, following standard usage in general relativity, $N$ is called the
lapse function, and $N^{A}$ the shift vector. We emphasize that the content
of this equation is simply that the time evolution of $\Sigma $ is into the
worldvolume $w$. Note that we can always chose a time evolution normal to
$
\Sigma $, {\it i.e.} take the shift vector to vanish, $N^{A}=0$ (see {\it 
e.g.} \cite{Hoppecan}).

At this point, we introduce the geometry of the worldvolume $w$. The
worldvolume can be represented in parametric form by the embedding functions 
$x^\mu = \chi^\mu (\xi^a)$, where $\xi^a = \{ t , u^A \} $ are local
coordinates for $w$, and $\chi^\mu$ the embedding functions ($a, b, \cdots
=
0, 1, \cdots , d$).

The  tangent vectors to $w$, $e_{a}:=e^{\mu }{}_{a}\partial _{\mu
}=(\partial _{a}\chi ^{\mu })\partial _{\mu }$, decompose in a part
tangential to $\Sigma $ and a part along the time evolution of $\Sigma $, 
\begin{equation}
e^{\mu }{}_{a}=\left( 
\begin{array}{l}
\dot{X}^{\mu } \\ 
\epsilon ^{\mu }{}_{A}
\end{array}
\right) \,.
\end{equation}
It follows that the lorentzian worldvolume induced metric, $\gamma
_{ab}:=g(e_{a},e_{b})$ , decomposes according to the familiar ADM expression
as, 
\begin{equation}
\gamma _{ab}=\left( 
\begin{array}{ll}
-N^{2}+N^{A}N^{B}h_{AB} & h_{AB}N^{B} \\ 
\,\,\,\,\,\,\,\,\,\,\,\,\,\,\,h_{AB}N^{B} &\,\,\,\, h_{AB}
\end{array}
\right) \,.
\end{equation}
Note that the worldvolume element is given by $\sqrt{-\gamma }=N%
\sqrt{h}$. The various geometrical quantities that characterize the
intrinsic geometry of $\Sigma $, such as its Riemann curvature etc., can be
decomposed by importing the appropriate expressions from the ADM treatment
of spacetime in canonical general relativity.

It is worth emphasizing the difference between, say, the lapse function in
this context versus the lapse function in canonical general relativity.
Whereas here it is a component of the velocity, in the latter case it is a
component of the metric.

The extrinsic curvature of the worldvolume $w$ along the $i$-th normal
vector field $\{n^{i}\}$ is defined by 
\begin{equation}
K_{ab}{}^{i}:=-g(n^{i},D_{a}e_{b})=K_{ba}^{i}\,,
\end{equation}
where $D_{a}=e^{\mu }{}_{a}\,D_{\mu }$ is the gradient along the vectors
tangential to $w$. It can be decomposed as 
\begin{equation}
K_{ab}{}^{i}=\left( 
\begin{array}{ll}
-n_{\mu }{}^{i}\ddot{X}^{\mu } &\,\, H_{A}{}^{i} \\ 
\,\,\,\,\,\,H_{A}{}^{i} & L_{AB}{}^{i}
\end{array}
\right) \,,
\end{equation}
where we have introduced the quantity 
\begin{equation}
H_{A}{}^{i}:=N\,\sigma _{A}{}^{0i}+N^{B}\,L_{AB}^{i}\,.
\end{equation}
We note that the time-time component of the extrinsic curvature is (minus)
the projection into the normals of the acceleration of $\Sigma _{t}$. For
the degenerate case of a relativistic particle, this is all there is. The
off-diagonal components involve the extrinsic twist of $\Sigma $. This is a
consequence of having broken the full normal rotation symmetry. The spatial
components involve the extrinsic curvature of $\Sigma $ along the normals 
$\{n^{i}\}$

The ADM decomposition of the mean extrinsic curvature of the worldvolume, 
$K^{i}:=\gamma ^{ab}K_{ab}{}^{i}$, is readily obtained as 
\begin{eqnarray}
K^{i} &=&N^{-2}[n_{\mu }{}^{i}\ddot{X}^{\mu }+2N^{A}H_{A}{}^{i}  \nonumber \\
&-&N^{A}N^{B}\,L_{AB}^{i}+N^{2}L^{i}]\,,  \label{eq:mean}
\end{eqnarray}
where $L^{i}=h^{AB}L_{AB}{}^{i}$ is the mean extrinsic curvature of $\Sigma $
along the normals to the worldvolume $\{n^{i}\}$. This is the generalization
to an extended object of the acceleration for a relativistic particle.

We can put the formalism we have developed to use in the analysis of the
dynamics of a relativistic extended object. For example, for a simple DNG
object, which extremizes the worldvolume, the equations of motion are given
by the vanishing of the mean extrinsic curvature, 
\begin{equation}
K^i = 0\,.  \label{eq:eom}
\end{equation}
If we specialize to normal time evolution, use of the expression (\ref
{eq:mean}) allows us to rewrite this equation in the form, 
\begin{equation}
n_{\mu}{}^i\ddot{X}^{\mu} + N^2 L^i = 0\,.  \label{eq:eom1}
\end{equation}
This identifies a part of the extrinsic curvature of $\Sigma $ as the
driving force in its dynamics. We emphasize that these expressions hold in
an arbitrary background spacetime.

The natural application of this geometrical approach is the Hamiltonian
formulation of a theory of relativistic extended objects. For simplicity, we
consider here only the case of a DNG object. Although the benefits of a
geometrical approach are already apparent in this case, the full power of
the formalism is displayed when considering higher-order curvature dependent
actions \cite{CGR2}, or additional worldvolume fields as for example in
the
case of superconducting membranes \cite{CR}.

We write the DNG action as $S=\int dtL[X,\dot{X}]$, where, with $\mu $ the
brane tension, the lagrangian is 
\begin{eqnarray}
L[X,\dot{X}] &=&-\mu \int_{\Sigma }\;d^{d}u\;\sqrt{-\gamma }  \nonumber \\
&=&-\mu \int_{\Sigma }\;d^{d}u\;N\;\sqrt{h}\,.
\end{eqnarray}
Therefore, the lapse function plays the role of the lagrangian density for a
DNG brane. The canonical momenta are given by $P_{\mu }:=\delta L/\delta 
\dot{X}^{\mu }$. Using the definition (\ref{eq:vel}) for the lapse function,
we find 
\begin{equation}
P_{\mu }=\mu \sqrt{h}\eta _{\mu }\,.  \label{eq:mom}
\end{equation}
The factor of $\sqrt{h}$ comes as no surprise -- momenta are densities. The
``shape'' functions of $\Sigma $ are the configuration variables, and their
conjugate momenta are proportional to the unit normal of $\Sigma $ into the
worldvolume. The phase space for our extended object is then naturally
associated with the geometry of $\Sigma $.

As expected from worldvolume reparametrization invariance, the hamiltonian
vanishes, $H[X,P]=\int d^{d}u\left[ P_{\mu }\dot{X}^{\mu }\right] -L[X,\dot{X%
}]=0$. However, according to the standard Dirac treatment of constrained
systems, the hamiltonian is a linear combination of the phase space
constraints that generate reparametrizations $\{{\cal C}_{0}, {\cal C}_{A}\}$, 
The explicit form of the constraints is easily obtained from the definition
of the momenta as 
\begin{eqnarray}
{\cal C}_A (X,P) &=& P_\mu \epsilon^\mu{}_A = 0\,,  \label{eq:cve} \\
{\cal C}_0 (X,P) &=& P^2 + \mu ^2 h = 0\,,  \label{eq:csc}
\end{eqnarray}
and the hamiltonian is
\begin{equation}
H[X,P]=\int_{\Sigma _{t}}[\lambda {\cal C}_{0}+\lambda ^{A}{\cal C}_{A}]\,,
\label{eq:ham}
\end{equation}
where  $\lambda$, $\lambda^A$ are lagrange multipliers.
Note that the lagrange multiplier $\lambda $ must be a scalar density of
weight $-1$ for the hamiltonian to be well-defined.

The first constraint, or vector constraint, is universal for all
reparametrizations invariant actions. Clearly it has no analogue in the case
of a relativistic particle, when $\Sigma$ degenerates to a point. It is easy
to see that it is the generator of spatial diffeomorphisms.

The second, or scalar constraint, is the generalization to extended objects
of the familiar hamiltonian  constraint for a parametrized massive
relativistic particle. It generates the evolution of $\Sigma$. For a
relativistic string, it is possible to exploit the triviality of its
one-dimensional $\Sigma$, by choosing proper length as the parameter along
the string, so that $h =1$, the potential is constant, and the scalar
constraint takes the form $P^2 + \mu ^2 = 0 $. This simple observation
explains the integrability of the string in a variety of background
spacetimes (see {\it e.g.} \cite{Bakas}). (It also strongly suggests that
the string may be integrable in {\it any} background spacetime.) Already for
a membrane ($d=2$), however, such a priviledged parametrization for $\Sigma$
is not available. The determinant of $h_{AB}$ contains four powers of
(spatial derivatives of) $X^\mu$, and the theory is highly non-linear, even
in a flat background. Hopes of integrability are evidently slim. Introducing
the bracket \cite{Hoppe,Nicolai} 
\begin{equation}
\{ X^\mu , X^\nu \} = \epsilon^{AB} \epsilon^\mu{}_A \epsilon^%
\nu{}_B \,,
\end{equation}
we can rewrite the scalar constraint for a relativistic membrane in the form 
\begin{equation}
P^2 +  \mu^2 \{ X^\mu , X^\nu \}^2 = 0\,.
\end{equation}
In this form we can make contact with current approaches to 
the canonical formulations of membrane dynamics 
(see  \cite{Nicolai,Smolin} ).
 
The constraints are first class. The Poisson algebra of the constraint
functions reproduce the familiar algebra of hypersurface deformation for $
\Sigma$, now seen as a hypersurface in the worldvolume $w$ \cite{Tei}. In
general it is not a Lie algebra since the structure ``constants" depend on
the configuration variables via the (densitized) metric
$h h^{AB}$ in the scalar-scalar Poisson
bracket. This is familiar from the constraint algebra of general relativity.
However, for a relativistic string, the triviality of the intrinsic geometry
of $\Sigma$ allows to set $h h^{AB} = 1 $, and we have a true Lie algebra,
which can be shown to be isomorphic to the algebra of the conformal group in
two dimensions \cite{KT}.

Let us consider the Hamilton equations. 
We have that
\begin{equation}
\dot{X}^\mu = 2 \lambda P^\mu + \lambda^A \epsilon^\mu{}_A \,.
\label{eq:ham1}
\end{equation}
It reproduces the definition of the momentum (\ref{eq:mom}), and it
identifies $\lambda^A = N^A$, $\lambda = N / 2 \mu \sqrt{h} $, assuming that 
$P_\mu$ is future-pointing. Using this information, the second Hamilton
equation is 
\begin{eqnarray}
\dot{P}_\mu &=& {\frac{1}{2}} h^{-1} \,\dot{h} \, P_\mu - N \mu \,\sqrt{h}
\,L^i\, n_{\mu \, i}   \nonumber \\
&+& \,\mu \, \sqrt{h} \, ( {\cal D}_A N ) \,\epsilon_\mu{}^A + {\cal D}_A (
N^A P_\mu )\,,
\end{eqnarray}
where, for convenience, we have left $\dot{h}$ in its implicit form. 
Note that $\dot{h} = 2 N h k$, where $k = h^{AB} L_{AB}{}^0$ 
is the mean extrinsic curvature of $\Sigma$
embedded as a hypersurface in $w$. For the analysis of this equation it is
convenient to take $N^A = 0$, and the shift constant ${\cal D}_A N = 0$. The
projection along the momentum is a mere identity, 
\begin{equation}
\dot{P}_\mu P^\mu = {\frac{1 }{2}} h^{-1} \, \dot{h} \, P^2\,,
\end{equation}
as can be seen using the scalar constraint, and its time derivative. The
normal projection 
\begin{equation}
\dot{P}_\mu n^\mu{}_i = - N \mu \,\sqrt{h}\,L_i\,,
\end{equation}
then reproduces the equations of motion in the form (\ref{eq:eom}).

This geometrical approach to the canonical formulation of 
relativistic extended  objects provides valuable insight into
their dynamics, and when it is relevant, into their canonical quantization. 
Furthermore, placing their
description on the same footing as canonical general relativity has the 
potential to reap mutual benefits, both technically and conceptually.

\end{document}